\documentclass[mathleft
]{an}
\usepackage{graphicx}
\usepackage{times}
\begin{document}

\Pagespan{789}{}
\Yearpublication{2006}%
\Yearsubmission{2005}%
\Month{11}%
\Volume{999}%
\Issue{88}%

\title{Precovery of near-Earth asteroids by a citizen-science project of the Spanish Virtual Observatory.}

\author{E. Solano\inst{1,2}\fnmsep\thanks{Corresponding author:
  \email{esm@cab.inta-csic.es}\newline}
\and C. Rodrigo \inst{1,2}
\and R. Pulido \inst{1,2}
\and B. Carry \inst{3}
}
\titlerunning{Using the Virtual Observatory to precover near-Earth asteroids}
\authorrunning{E. Solano, C. Rodrigo, R. Pulido \& B. Carry}
\institute{Centro de Astrobiolog\'{\i}a (INTA-CSIC), Departamento de Astrof\'{\i}sica. Campus Villafranca. P.O. Box 78, E-28691 
Villanueva de la Ca\~{n}ada, Madrid, Spain
\and 
Spanish Virtual Observatory
\and
IMCCE, Observatoire de Paris, UPMC, CNRS, 77 Av. Denfert Rochereau 75014 Paris, France
}

\received{xx Dec 2012}
\accepted{xxxxx}
\publonline{later}


\abstract{%
 This article describes a citizen-science project conducted by the Spanish Virtual Observatory (SVO) to improve the orbits of near-Earth 
asteroids (NEAs) using data from astronomical archives. The list of NEAs maintained at the Minor Planet Center (MPC) 
is checked daily to identify new objects or changes in the 
orbital parameters of already catalogued objects. Using NEODyS we 
compute the position and magnitude of these objects at the observing epochs of the 938 046 images comprising the 
Eigth Data Release of the Sloan Digitised Sky Survey (SDSS). If the object lies within the image boundaries and the 
magnitude is brighter than the limiting magnitude, then the associated image is visually inspected by the project's 
collaborators ({\it the citizens}) to confirm or discard the presence of the NEA. If confirmed, accurate coordinates 
and, sometimes,  magnitudes are submitted to the MPC. Using this methodology, 3,226 registered users have made during the first 
fifteen months of the project more than 167,000 
measurements which have improved the orbital elements of 551 NEAs 
(6\% of the total number of this type of asteroids). Even more remarkable is the fact that these results have been 
obtained at zero cost to telescope time as NEAs were serendipitously observed while the survey was being 
carried out. This demonstrates the enormous scientific potential hidden in astronomical archives.
\\
 The great reception of the project as well as the results obtained makes it a 
valuable and reliable tool for improving the orbital parameters of near-Earth asteroids.}

\maketitle

\section{Introduction}
Small Solar System bodies were defined in 2006 by the IAU as those objects that are neither planets nor dwarf 
planets, nor satellites of a planet or dwarf 
planet\footnote{http://www.iau.org/static/resolutions/Resolution\_GA26-5-6.pdf}. They are remnants of the early stages of the solar
 system 
formation process and the vast majority of them can be found in two distinct areas: the asteroid belt and the 
Edgeworth-Kuiper belt. Nevertheless, due to the Yarkovsky effect, some main-belt asteroids can be 
thermally drifted from their source locations to chaotic resonance zones capable of injecting them into the inner Solar System 
(Bottke et al. 2006). 

Near-Earth asteroids (NEAs) are small bodies whose orbits bring them into close proximity with the Earth (perihelion
 distance q $<$ 1.3 AU)\footnote{http://neo.jpl.nasa.gov/neo/groups.html}. Potentially 
Hazardous Asteroids (PHAs) are defined based on parameters that measure the object's potential to make threatening 
close 
approaches to the Earth. NEAs having a minimum orbital intersection distance (MOID) $<$ 0.05 AU and an absolute 
magnitudes H $<$ 22 (which corresponds to objects larger than about 100\,$\mathrm{m}$) are considered 
PHAs. 

Although the first NEA was not discovered until 1898 (Eros) and the first NEA that actually crosses Earth's orbit 
(Apollo) was not found until 1932, already in the 17th century Edmond Halley raised the danger of a potential impact 
with the Earth (Chapman 2004). The discovery in the late 1980s of NEAs passing by the Earth at distances 
comparable to that of the Moon or the impact of the comet Shoemaker-Levy 9 to Jupiter in July 1994 led to an 
increased awareness of the potential threat of these objects. In 1998 NASA set the goal of discovering within one 
decade 90\% of NEAs with diameters greater than $>$ 1\,$\mathrm{km}$. This objective was extended in 
2005 to detect, track, catalogue  and characterize the physical properties of 90\% of NEAs equal to or greater than 
140 meters in diameter before 2020. The projects that came up to fulfil these mandates gave rise to a significant 
increment in the number of NEAs. A detailed description of the main NEA discovery programmes can be found in 
Vaduvescu et al. (2011). The resolution approved in the IAU General Assembly held in Beijing in August 2012 to 
{\it ``coordinate and collaborate on the establishment of an International near-Earth objects (NEO) early warning 
system, relying on the 
scientific and technical advice of the relevant astronomical community, whose main purpose is the reliable
identification of potential NEO collisions with the Earth, and the communication of the relevant parameters to 
suitable decisions makers of the nation(s) involved''}, is a significant step further to highlight the importance of 
having an accurate knowledge of the number, size and orbital behaviour of the small objects that populate the Earth's
vicinity. 

Discovery alone is not enough to quantify the thread level of a NEA. Above all, it is 
necessary to compute reliable orbits through accurate astrometric positions covering a period of time as long as 
possible. This can be achieved using two complementary approaches: performing follow-up observations after discovery 
or mining astronomical archives with information (images, catalogues) prior to the discovery date. Precovery (short 
for ``pre-discovery recovery") is the term that describes the process of identifying an object in archive data whose 
presence was not detected at the time the observation was made. Although different valuable results obtained by 
precovery searches can be found in the literature (e.g., the improvements of the orbital solutions of the Hale-Boop 
(McNaught \& Cass 1995) and Swift-Tuttle comets (Marsden et al. 1992)), the importance of this approach was not
 clearly proved until the discovery of Apophis in 
December 2004. Follow-up observations performed within one week after its discovery rose Apophis to an unprecedented 
level of hazard alert in the Torino Scale, which describes and quantifies the risks associated with asteroid 
collisions according to the probability of impact and size of the impactor (Binzel 2000). The level of 4 reached by 
Apophis was due to its non-negligible probability of 2.6\% to hit the Earth in 2029 and to cause localized 
destruction: with a diameter of about 300\,$\mathrm{m}$, Apophis would have created an impact crater of 
about 3\,$\mathrm{km}$.  
It was not until the use of precovery positions taken in March 2004 (nine months before discovery) 
when the impact could be ruled out. In parallel to an accurate determination of the risk of impact, extending the 
observed arc of the orbit through precovery observations also allows other types of studies like, for instance, 
orbital evolution (Asher \& Steel 1993).

Despite its potential, very few projects have been devoted to look for asteroids in astronomical archives. Some
 pioneering 
initiatives started in the nineties using photographic plates: 
AANEAS\footnote{http://msowww.anu.edu.au/$\sim$rmn/aaneas.htm}, 
AANEOPP\footnote{http://www.arcetri.astro.it/science/aneopp/} or DANEOPS\footnote{http://earn.dlr.de/daneops/}, 
to name a few. Taking advantage of the long exposures of the photographic plates, NEAs 
could be efficiently
 discriminated from the background stars by their long trailed appearance. While photographic archives represent 
excellent resources to expand the timeline of observations to decades instead of months or years, accessibility
 problems, modest limiting magnitudes and, in some cases, bad state of conservation strongly limit their real usage.
 Moreover, visual inspection of plates is a very time-consuming task that requires an enormous amount of work. 

Digitisation represented an important leap forward in order to efficiently exploit the scientific potential hidden 
in astronomical archives. Information in digital format can be analysed by computers and can be accessible 
remotely. Nevertheless, the high degree of heterogeneity and the lack of interoperability
among astronomical services (different access procedures, data representation, retrieval methods,...) hinders the 
scientific analysis of combined information from more than one astronomical resource and just very few projects
 (e.g., Boattini et al. 2001, Vaduvescu 2011 and references therein) have been devoted so far to mine
 astronomical archives. These drawbacks can be overcome if the discovery, gathering and analysis of the information 
is carried out in the framework of the 
Virtual 
Observatory\footnote{http://www.ivoa.net}, an international initiative designed to provide the
astronomical community with the data access and the research
tools necessary to enable the exploration of the digital, multiwavelength
universe resident in the astronomical data archives. 

Every single image taken by the most important ground and space-based astronomical observatories eventually end up in 
open archives, freely available on the web. This represents an immensely data-rich field where the general public can
significantly contribute, in particular in projects related to classification, pattern recognition and outlier 
identification where the visual inspection has proved exceptionally good. Although the huge amount of data available 
in astronomical archives and services make visual analysis by a single individual or reduced group of collaborators 
impossible, this problem can be overcome if a crowdsourcing scenario is considered. This is the 
basis of the citizen-science projects whose most popular examples in Astrophysics are Galaxy 
Zoo\footnote{http://www.galaxyzoo.org} and its extension to Zooniverse\footnote{http://www.zooniverse.org}. The
discovery of extrasolar planet candidates using the Kepler Public Archive (Fischer et al. 2012) or the identification 
of a new class of compact star-forming galaxies ({\it The Green Peas}, Cardamone et al. 2009) are excellent 
examples of the ability of human classifiers to recognize unusual patterns which computer search algorithms are 
unable to spot.
   
In this paper we describe a citizen-science project designed by the Spanish Virtual 
Observatory\footnote{http://svo.cab.inta-csic.es} (SVO) to precover NEAs in the Data Release 8 of the Sloan Digital 
Sky Survey (SDSS, Hiroaki et al. 2011). Through 
visual inspection of sequences of images, the user is requested to identify the asteroid and measure its coordinates. 
After passing a number of quality checks, the asteroid positions are sent to the Minor Planet 
Center\footnote{http://minorplanetcenter.net} (MPC) to improve the associated orbital parameters. The project gives the 
public the opportunity to participate in an attractive initiative going through the same steps as professional 
astronomers (data acquisition, data analysis and publication of results) and making useful contributions to a better 
knowledge of 
potential threads of collision with the Earth. The public release of the system took place 
on July 2011, and after fifteen months, more than 3000 users have participated in it. 

In this paper we describe the layout of the system and the results obtained after one year of operations.

\section{The system}

The system\footnote{http://www.laeff.cab.inta-csic.es/projects/near/main/?\&newlang=eng} is a PHP-based 
application whose 
primary function is to provide SDSS images to the users and to gather the astrometric measurements they made. The 
main components and functionalities of the system are as follows:

\subsection{The survey}

At present, there is only one archive publicly available through the system: the Eight Release of the Sloan 
Digitised Sky Survey (SDSS/DR8). SDSS (York et al. 2000) is one of the most important surveys in the history of 
astronomy and has obtained, over twelve years of operations, almost a million images covering more than 14,500 
square degrees (more than a quarter of the sky).

SDSS used a dedicated 2.5\,$\mathrm{m}$ telescope at Apache Point Observatory (New Mexico, observatory code 645), 
equipped with a 
large-format mosaic CCD camera to image the sky in five optical bands: $\mathrm{r}, \mathrm{i}, \mathrm{u}, 
\mathrm{z}$, and $\mathrm{g}$ (Fukugita et al. 1996). The
 120-megapixel camera 
has a pixel scale of 0.4\arcsec $ $ covering 1.5 square degrees of sky at a time. The imaging survey is taken in 
drift-scan,
 i.e., the camera continually sweeps the sky in strips along great circles. A specific scan is identified with the 
{\bf run} number. Each strip consists of six parallel scanlines (identified by the camera column or {\bf camcol} 
number), 13\arcmin $ $ wide. The scanlines are divided into {\bf fields} which constitute the fundamental units of SDSS
images. An additional number, {\bf rerun}, specifies how the image was processed. 

In the SDSS observing sequence, a given 
point on the sky passes through the five filters in succession.The effective integration time per filter is 
54.1\,$\mathrm{s}$, 
and the time for passage over the entire photometric array is about 5.7 minutes. The survey is complete for point 
sources to limiting magnitudes 22.0, 22.2, 22.2, 21.3 and 20.5 (Juri\'{c} et al. 2002). Astrometric positions are 
accurate to 0.1\arcsec $ $ for sources brighter than 20.5 mag (Pier et al. 2003).

Large sky coverage and field of view, faint limiting magnitude, excellent astrometric and accurate multicolor 
photometry make SDSS an excellent resource to identify, discover and characterize minor bodies in the solar system. 
(e.g., Ivezi\'{c} et al. 2001). The result of this is the SDSS Moving Object 
Catalog\footnote{http://www.astro.washington.edu/users/ivezic/sdssmoc/sdssmoc.html} containing data for more than 
470,000 moving objects observed prior to March 2007. 

Nevertheless, despite the high performance of the SDSS photometric pipeline to detect asteroids, there exist some 
limitations in the identification process that affect its completeness (i.e., the fraction of moving objects 
recognized as such by the moving object algorithm). So, for instance, objects moving faster than 0.5 deg/day may be 
classified as 
separate objects. Objects with lower velocities, on the other hand, require to be detected in at least three bands 
to fit the two components of the proper motion and have a total motion significantly different from zero (2$\sigma$)
 for them to be declared moving. Also, asteroids in SDSS images taken after 
the latest release of the Moving Object Catalogue (March 2007) remain yet unidentified.

For every single image in the SDSS/DR8, we harvested the necessary metadata to characterize its temporal and spatial 
boundaries. In particular, the following parameters were ingested in our database: image identifiers 
(run,rerun,camcol,field), mid-exposure time, right ascension and declination of the image center as well as 
image size. 

\subsection{NEA's basic information retrieval }

The gathering of this information is carried out in two steps:

\begin{itemize}
 \item [$\bullet$] Bulk download: The list of NEAs was obtained from the Minor Planet Center. In addition to the name and/or 
provisional designation we gather 
information on the orbital elements, the minimum distance between the orbit of the Earth and the minor planet, the 
absolute visual magnitude and the number of oppositions (or the arc length in days if there is only one opposition) 
at which the object has been observed. 

Once this information has been retrieved, we use NEODyS\footnote{http://newton.dm.unipi.it/neodys2/} to obtain, for 
every NEA, the coordinates (RA, DEC) and the magnitude (V) at different epochs. A one-day step is assumed and the total 
time range covered is given by the first and last image in the SDSS/DR8. The asteroid coordinates at the image observing time 
are calculated by linear interpolation. Images containing a NEA brighter than the 
limiting magnitude (set to V=22) are downloaded from the SDSS archive to our system. \\

\item [$\bullet$] Daily update: On a daily basis, the system checks the Minor Planet Center for new NEAs or already existing NEAs that 
have changed their orbital parameters. For these objects, new ephemeris are calculated using NEODyS and the 
whole process is repeated. 
\end{itemize}

The results presented in this paper are based on the status of the MPC database on 2012 September 18. 

\subsection{The measurement process}

\begin{figure*}
\includegraphics[width=17cm]{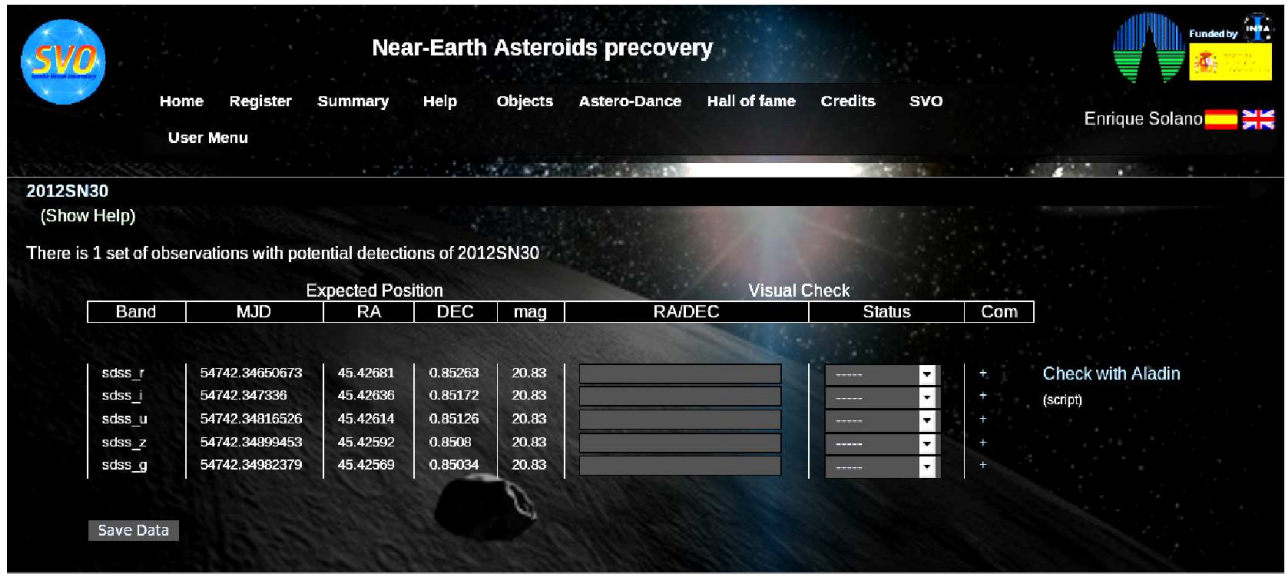}
\caption{System's input window. For a given asteroid (2012 SN30) a block with five consecutive SDSS images is shown. 
The observing epoch of each image as well as
the expected position and V magnitude of the asteroid as provided by NEODyS is given in columns 2-5. Accurate 
positions of the asteroid measured by the participants and the associated status ({\it confirmed, out of field, too 
faint}) can be included in columns 6-7. Column 8 allows the user to include a comment about the measurements. Images 
are displayed by clicking on {\it ``Check with Aladin''} (see Figure \ref{figure2}). The measurements are stored in the 
system clicking on the 
{\it ``Save Data''} button.}
\label{figure1}
\end{figure*}

\begin{figure*}
\includegraphics[width=17cm]{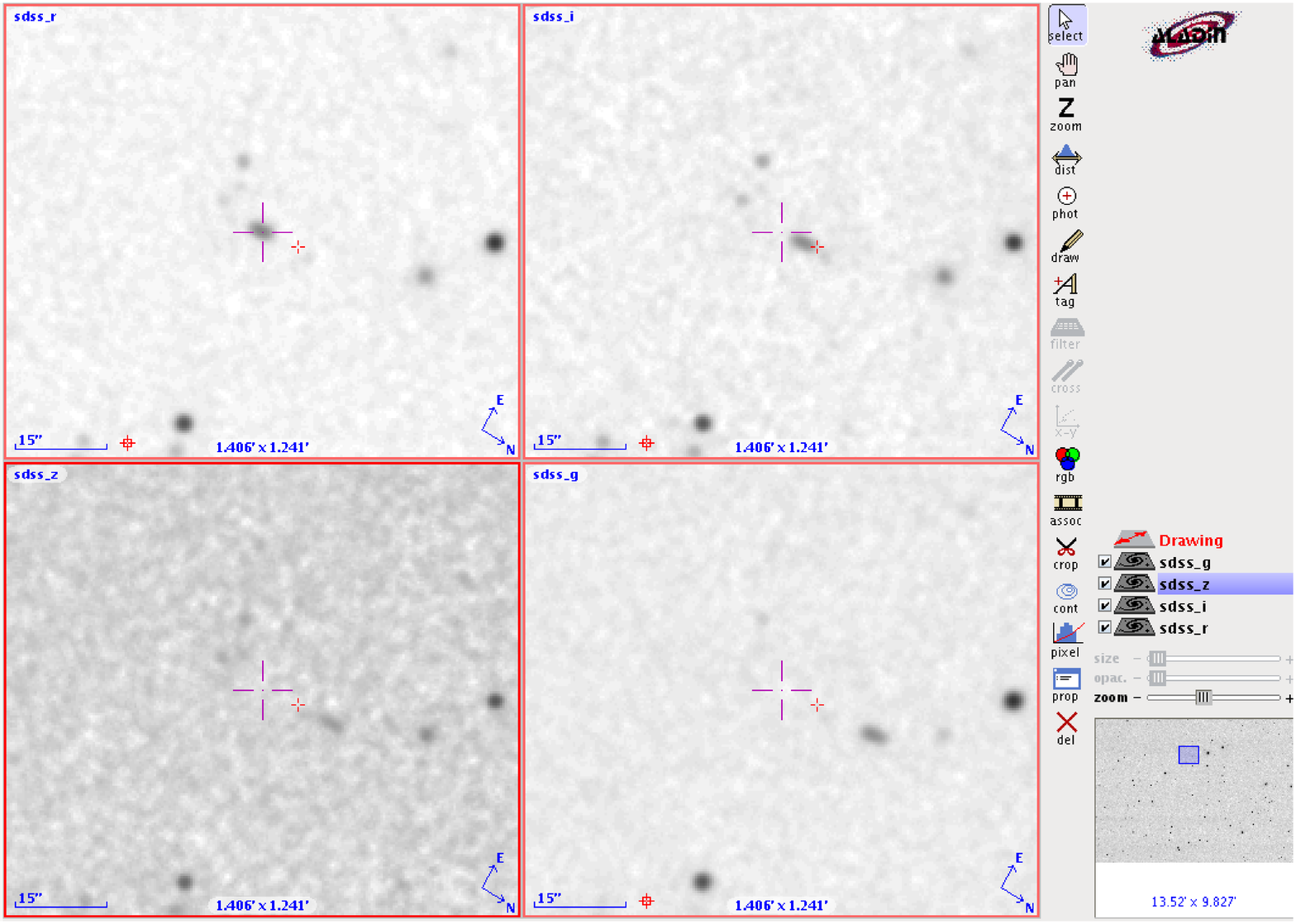}
\caption{Asteroid identification. The asteroid 2007JZ20 is clearly seen moving in the sequence of SDSS images from 
South to North. The small red cross indicates the expected position as computed by NEODyS. The user must put the 
large magenta cross on the asteroid and paste the coordinates in the table shown in Figure \ref{figure1}.}
\label{figure2}
\end{figure*}

\begin{itemize}
\item [$\bullet$] Registration: People willing to participate in the project must register first. Some basic 
information (name, e-mail address, country, birth year, the channel through which they knew the project
 and institution (if any)) is requested for statistical purposes only. Name is also necessary for the information to 
be sent to the Minor Planet Center. The new user is informed by e-mail when the authorization request is 
accepted which allows him/her to start with the measurement process. \\

\item [$\bullet$] Object selection: The system provides the user with a list of NEAs. By clicking 
on the asteroid name the system returns a table with a list of images that include, for each of them, the observing 
time and the expected position and V magnitude of the asteroid as calculated by NEODyS (Figure \ref{figure1}). Images with the 
same run and camcol are grouped based on the observing time and the filter sequence ($\mathrm{riuzg}$). Each group of images 
can be visualized using Aladin\footnote{http://aladin.u-strasbg.fr/} (Bonnarel et al. 2000). Aladin 
is a VO-compliant interactive software sky atlas allowing to visualize digitized astronomical images, 
superimpose entries from astronomical catalogues or databases, and interactively access related data and 
information from VO services. \\

\item [$\bullet$] Identification: The procedure of identifying NEAs in SDSS images relies on the visual comparison of 
images of 
the same region of the sky taken at different moments. The vast majority of the objects recorded in the images are 
stars and galaxies that will appear in the same position in all the images. On the contrary, NEAs are nearby objects
 with proper motions of several arcseconds per minute and they will appear in slightly different positions in the 
roughly 5 minute sequence of $\mathrm{riuzg}$ images. To help users in the identification,
the predicted position of the asteroid as given by NEODyS is marked with a red cross in the screen 
(Figure \ref{figure2}). The 
pre-defined Aladin setting shows a multiview of SDSS images ordered in time with a 4x zoom level. Also, a logaritmic 
transfer function was adopted to provide an optimal 
representation of the dynamic range and optimum image contrast. Anyway, collaborators can take advantage of all Aladin 
capabilities (zoom and contrast level, RGB composite image, movie 
generation, ..) to help themselves in the asteroid identification. 

There are two main requirements to confirm the asteroid identification: the object must be located close to the 
predicted position, and the apparent motion (both in Right Ascension and Declination) must be consistent with the 
temporal sequence of the SDSS images.  

Once identified, the user must measure the position of the asteroid in the different images and include it 
in the RA/DEC column (Figure \ref{figure1}). After this, he/she must choose among the three values available in the 
column
``status", namely: {\it Confirmed}, {\it out of field} (if the asteroid lies beyond the image's boundaries) or 
{\it too faint} (if the 
expected position of the asteroid is within the image boundaries but the user has not been able to identify it). The 
users has also the option of including a comment 
providing additional information about the measurement he/she has made. Finally, the measurement is recorded by 
clicking ``Save Data''. Users can only classify a given image once.

The position of every single asteroid is measured, at least, by fifteen different users. If the standard deviation
 is lower than 1\arcsec $ $ both in RA and DEC, the asteroid is removed from the list and the consensus coordinates made available to
 the SVO staff in a private website for a final visual inspection before submitting the information to the Minor 
Planet Center. If the astrometric errors remain above the threshold after having been measured by 45 different 
users, the asteroid is also removed from the list and further analyzed by our group. 

The system includes an on-line step-by-step tutorial to provide guidance and training to the users. A Helpdesk which 
guarantees a direct contact with the SVO staff is also implemented. 

\end{itemize}

\section{Results}

\begin{figure}
\includegraphics[width=8cm]{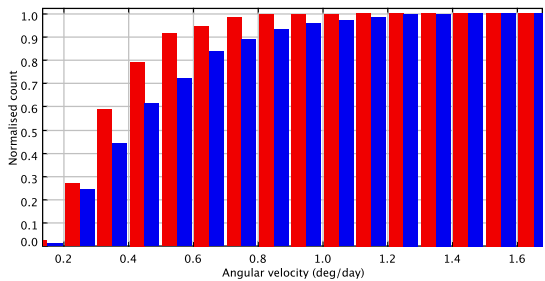}
\caption{Cumulative histogram showing the distribution of angular velocities for NEAs included in the MOC (red) and those 
identified in this work and, therefore, not included in the MOC (blue). We see how 90\% of the NEAs included in the MOC show angular 
velocities smaller than 0.6 deg/day whereas this percentage is only 70\% for the NEAs identified in our project.}
\label{MOC}
\end{figure}

At the time of writing (September 2012), over 3200 people have logged in to the system. 15\% of the users measured 
at least 
one asteroid with 1\% contributing the top 50\% of the total measurements. These are typical values for online projects 
(see, for instance, Schwamb et al. 2012 and references therein). 2351 new measurements (position and, sometimes, magnitudes) 
corresponding to 551 NEAs ($\approx$ 6\% of the 
the total census of NEAs) have been reported to the Minor 
Planet Center. 
 These measurements have been published in different issues of the Minor Planet Circulars available from the 
SAO/NASA Astrophysics Data System\footnote{http://www.adsabs.harvard.edu/}. It is important 
to remark that 1696 of our measurements (73 \%) were made on images taken before May 2007 and, therefore, correspond 
to 
asteroids not identified in the SDSS Moving Object Catalogue 
(MOC)\footnote{http://www.astro.washington.edu/users/ivezic/sdssmoc/sdssmoc.html}. The apparent angular velocity 
of these NEAs was compared to that of the NEAs included in the MOC (Figure \ref{MOC}). As expected from the MOC data 
selection criteria, it misses a significant percentage of NEAs having velocities larger than 0.5 deg/day. 

The top-five nations visiting the project have been Spain (63.3 \%), Mexico (6.7 \%), Argentina (5.1 \%), 
Venezuela (4.2 \%) and Colombia (3.8 \%). The remaining 16.9 \% of the visitors came from 16 other countries. 
Although the site is offered both in Spanish and English, it is mainly accessed by Spanish-speaking 
countries. Age distribution, on the contrary, is clearly less abrupt indicating the attractiveness of the 
project for people of very different ages. 

Regarding the way how participants got to know the project, although channels like social networks and/or blogging 
play a non-negligible role, the publication in national newspapers has been, by far, the most successful form to 
contact potential participants. 

\begin{figure}
\includegraphics[width=8cm]{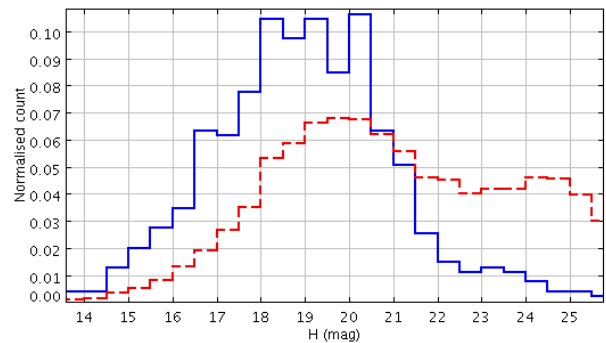}
\caption{Distribution of the absolute magnitude H for all NEAs catalogued by the MPC (red, dashed line) and those
measured in our project (blue, solid line).}
\label{Hmag}
\end{figure}

In Figure \ref{Hmag} we compare the normalised distribution of the absolute magnitude H as provided by the MPC for 
all the 
known NEAs and for those identified in our project. These range 
from 13.8 to 25.5 with a maximum at H $\sim$ 19--20, value that can be regarded as the completion limit for our
project. Assuming an albedo (defined as the ratio between the intensity of the radiation reflected from a surface 
and the incident radiation) of 23\%, typical of the S-type asteroids which represent the predominant group in the
inner part of the Solar System, this limit in magnitude translates into a minimum diameter of 
$\sim$ 300\,$\mathrm{m}$ following 
the equation (Yoshida \& Nakamura 2007 and references therein)

\begin{equation}
\log (D) = 3.1295 - 0.5 \log (p) -0.2 H 
\end{equation}

where D is the asteroid diameter in $\mathrm{km}$, $\mathrm{p}$ is the albedo and $\mathrm{H}$ is the absolute 
magnitude. The faintest H magnitude 
reached in our project (H=25.5) would correspond to a diameter of $\sim$ 25\,$\mathrm{m}$. 

\begin{figure}
\includegraphics[width=8cm]{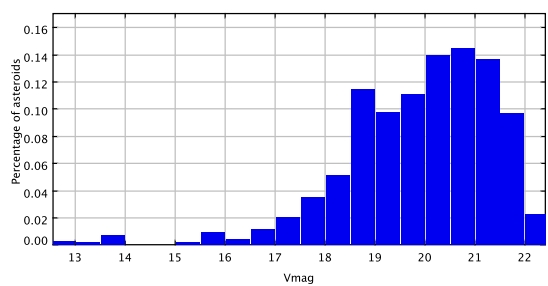}
\caption{Distribution of V apparent magnitudes for the NEAs analyzed in this paper.}  
\label{Vmag}
\end{figure}

Figure \ref{Vmag} plots an histogram showing the apparent magnitude V as provided by NEODyS for the 551 NEAs measured in the
project. A limiting magnitude V: 21--21.5 is reached, in good agreement with the nominal SDSS limiting magnitude. 

Regarding the improvements of the orbital accuracy, the main results have been:

\begin{itemize}
 \item [$\bullet$] {\it Precovery:} We have found precovery observations, that is, prior to the first observation 
reported in the MPC database for  
130 NEAs (24 of them classified as PHAs). This represents 23\% of the total sample of NEAs reported to the MPC. 
Thirty-one, eighteen and two (2010 FC6, 2011 MB2) asteroids have extended their arc lengths more than one thousand, 
two thousand and four thousand days, respectively. We have also augmented by more than a factor of two the arc 
lengths of three asteroids (1999 GR6,  2004 FY31, 2005 FC) whose previous temporal coverage was less than 15 days. 
In all these cases 
our astrometry significantly improved the orbital elements. For the rest of 
precoveries, the temporal coverage was extended from a few days to less than three years. \\

\item [$\bullet$]  {\it Arc extension:} We provided measurements after the last observation 
given in MPC for 29 NEAs (4 PHAs). For two of 
them a new opposition was added. The two most remarkable results were found for 1999 VN6 (Amor) whose arc length was 
increased almost two years (617 days) and for 2009 CP5 (Apollo) with an increase in the arc length from 34 to 245 
days.  \\

\item [$\bullet$]  {\it Intermediate oppositions:} New intermediate opposition (being our 
measurements the only data available at those oppositions) were found for 24 NEAs (3 PHAs). Of particular interest 
is 2010 JG123 (Apollo) for which the number of oppositions increased from 4 to 7. \\

\item [$\bullet$]  {\it Single oppositions:} We added new intermediate observations for 56 (11 PHAs) 
single-opposition asteroids. Remarkable is the case of 2001 UO27 (Amor), with only 16 observations and for which we
added 4 new measurements. \\

\item [$\bullet$] {\it Potential spacecraft targets:} After the success of missions like 
NEAR-Shoemaker\footnote{http://near.jhuapl.edu/} or 
Hayabusa\footnote{http://www.isas.jaxa.jp/e/enterp/missions/hayabusa/index.shtml}, new space missions to near-Earth
asteroids are being planned at all major space agencies. Finding a suitable asteroid is one of the most 
critical aspects of these projects and the final selection is tightly linked to the optimization of mission 
costs, in particular the required amount of propellant. $\Delta \nu$, a measurement of the changes in speed made by 
the propulsion system during the mission can be used as an indicator of such costs. Asteroids with 
$\Delta \nu$ $<$ 6--7 km/s fall into the category of possible destinations for human exploration. 

We improved the orbit of 68 of all 4298 (as of September 2012) potential mission targets with 
$\Delta$ $\nu$ $<$ 7 km/s. The major characteristics (type, H magnitude, arc length and close approaches before 
2050) of the asteroids having $\Delta$ $\nu$ $<$ 6 ({\it ``cheap targets''}) are summarised in Table \ref{table2}. 
It is 
worth noting 2009 EK1 and 2006 KM103, whose arc lengths were dramatically increased thanks to our observations. 
Almost 76\% of the targets (14/19) in Table \ref{table2} have observational arcs longer than 0.1 years 
(a rough measure of how well the
orbit is known (Mainzer et al. 2011)). Moreover, 42\% (8/19) have absolute magnitudes lower than H=22.0 which would 
correspond, in a first approximation, to a diameter of 100\, $\mathrm{m}$, appropriate for the landing of a 
spacecraft. 

\end{itemize}

A summary of the results describe above is given in Tables \ref{table1} and \ref{table2}. For the rest of 312 NEAs (551 reported to 
MPC minus 239 included in Table 1) we added measurements at intermediate positions. 

\begin{table}
 \caption{Summary of the measurements reported to MPC. (1): Extended arc beyond last observation reported by the 
MPC. (2): New intermediate positions. (3): New observations for single opposition asteroids. }
\label{table1}
\centering
\begin{tabular}{r r r r r r}
 \hline\hline
                                         & PHA   & Apollo   & Aten & Amor     & Total\\
Precovery                                & 24    &    42    &  6   &  58      & 130  \\
Extended$^{(1)}$                         &  4    &    13    &  ---  &   12     & 29   \\
Intermediate$^{(2)}$                     &  3    &    6     &  --- &  15      & 24   \\
Single$^{(3)}$                           &  11   &   15     &  1   &  29      & 56     \\
Total                                    & 42    &   76     &  7   & 114      & 239 \\
\hline
\end{tabular}

\end{table}

\begin{table*}
\caption{Asteroids with $\Delta \nu <$ 6 analysed in our project }
\label{table2}
\centering
\begin{tabular}{l l l l r c l}
 \hline\hline
Name       & Type   & $\Delta \nu$ & H mag. & Arc length & Close app.  & Comments\\
           &        &                &        & (days)     & ($<$2050) & \\
2005 EZ169 & PHA    & 4.49           & 25.1   &   23      & 0 & Precovery. Orbit extended from 20 to 23 days\\
2005 HB4   & PHA    & 4.62           & 24.3   &   77      & 1 & Precovery. Orbit extended from 49 to 77 days\\
2003 BN4   & PHA    & 4.75	     & 24.6   &   84      & 2 & Precovery. Orbit extended from 58 to 84 days\\
1996 FO3   & PHA    & 4.90           & 20.4   &  5990     & 2 & New intermediate opposition\\
2006 UN    & Amor   & 4.93           & 25.5   &  56       & 0 & New observations for single-opposition asteroid\\
2006 YF    & Apollo & 4.98           & 20.9   &  92       & 4 & New observations for single-opposition asteroid\\
2000 SL10  & PHA    & 5.08           & 21.8   &  69       & 1 & New observations for single-opposition asteroid\\
2005 SQ9   & Apollo & 5.14           & 23.0   &  29       & 1 & New observations for single-opposition asteroid\\
2009 EK1   & PHA    & 5.19           & 21.3   & 4006      & 2 & Precovery. Orbit extended from 19 to 2650 days\\
2004 UR    & PHA    & 5.33           & 23.0   &   49      & 1 & Precovery. Orbit extended from 34 to 49 days\\
2009 UA    & Amor   & 5.46           & 23.7   & 10        & 0 & New observations for single-opposition asteroid\\
2005 LD    & Apollo & 5.51           & 23.0   & 186       & 1 & New observations for single-opposition asteroid\\
2000 CM33  & PHA    & 5.60           & 21.2   & 1259      & 2 & Orbit extended from 1227 to 1259 days\\
2010 VA1   & PHA    & 5.66           & 20.0   & 4071      & 2 & Precovery. Orbit extended from 4043 to 4071 days\\
2009 TV4   & Amor   & 5.67           & 23.3   & 10        & 0 & New observations for single-opposition asteroid\\
2006 KM103 & PHA    & 5.78           & 20.0   & 4594      & 2 & Precovery. Orbit extended from 1804 to 3868 days\\
2008 UD1   & PHA    & 5.84           & 19.3   & 1388      & 2 & Precovery. Orbit extended from 29 to 47 days\\
2001 DS8   & Amor   & 5.96           & 22.7   & 60        & 0 & New observations for single-opposition asteroid\\
2008 UT2   & Apollo & 5.98           & 25.2   & 6         & 0 & New observations for single-opposition asteroid\\
\hline
\end{tabular}
\end{table*}

\section{Future work}

Our main goal in the mid-term is to expand the capabilities of the system by including new surveys, 
new functionalities and new types of asteroids. In particular, the following improvements are foreseen:  

\subsection{New type of asteroids}
\begin{itemize}
 \item [$\bullet$] Mars crossers: A Mars-crosser is an asteroid whose orbit crosses that of Mars. 8987 objects are classified as 
Mars crossers by the Minor Planet Center and a subset of them (those numbered, with more than a single opposition and
with ephemeris available from Astdys\footnote{http://hamilton.dm.unipi.it/astdys/}) will be included in our system 
in the next data release.
\end{itemize}

\subsection{New surveys}

\begin{itemize}
\item [$\bullet$] UKIDSS\footnote{http://www.ukidss.org}: Considered the successor to 2MASS, the UKIRT InfraRed Data Sky Survey 
(UKIDSS, Lawrence et al. 2007) began in May 2005 and will cover 7500 square degrees of the Northern sky, extending 
over both high and low Galactic latitudes to K$_{s}$=18.3. UKIDSS is considered the near-infrared counterpart to the Sloan 
survey. The Eight Data Release\footnote{http://surveys.roe.ac.uk/wsa/} 
is publicly available since April 2012 and the images are accessible through Virtual Observatory access protocols 
(SIAP).  

Assuming for NEAs a typical color of 1.5 $\leq$ (V-K) $\leq$ 2 (Baudrand et al. 2001), the limiting magnitude would
 be V $\sim$ 20 \\

\item [$\bullet$] VISTA\footnote{http://www.vista.ac.uk/index.html}: VISTA is a 4-$\mathrm{m}$ class wide--field survey telescope located in
 Chile and 
operating in the infrared. Since 2010 and for five years VISTA will be conducting six surveys with different area
 coverage, depth and scientific goals. The first public data releases\footnote{http://horus.roe.ac.uk/vsa/index.html}
 of the different surveys are already available through VO tools. VHS and VIKING are the two VISTA surveys with the 
largest area coverage and with limiting magnitudes of K$_{s}$=20.0 and K$_{s}$=21.2 respectively.

\end{itemize}

\subsection{New functionalities}

\begin{itemize}
 \item [$\bullet$] Rapid response: The number of astronomical transient detections will grow enormously over the next few years 
with missions like Gaia or LSST. Newly discovered asteroids of special interest should not wait until the normal 
measurement process ends (typically of the order of weeks) as rapid identification in archives can be the key for a 
reliable characterization. We plan to implement in our system a rapid response mode able of precovering asteroids as 
soon as their ephemeris are available in NEODys (typically 24-48 hours after discovery). 
\end{itemize}

\section{Conclusions}
The increasing number of digital astronomical databases available via the Internet and hosting terabytes of data 
over the electromagnetic spectrum as well as the new possibilities that the Virtual Observatory offers in terms of 
data discovery, gathering and analysis have made archives a fundamental tool in modern astrophysics. Archive-based 
projects transcend the boundaries of the professional astronomy and opens a world of opportunities in the fields of 
education and outreach.

We have described here the main aspects of an outreach programme conducted by the Spanish Virtual Observatory. 
The project enlists members of the general public to visually identify near-Earth asteroids in archive images. The 
Web site was launched on 2011 July, and after fifteen months, more than 3,200 users have made more than 167,000 
measurements over 551 NEAs (6 \% of the total census) that were used by the Minor Planet Center to improve their 
orbits.The fact that a large percentage of objects 
were unnoticed for the automated detection mechanisms implemented in the SDSS pipeline query stresses the validity, 
complementarity and powerfulness of our citizen-science project. 


\acknowledgements
This publication has been made possible by the participation of more than 3,200 volunteers without whom this work
 would not have been possible.

 The SVO-NEA system has been developed under the Spanish Virtual Observatory project (Centro de Astrobiolog\'{\i}a, 
INTA-CSIC)  supported from the Spanish MICINN through grants AyA2008-02156 and AyA2011-24052. This research has made 
use of Aladin and NEODyS. 

Funding for SDSS-III has been provided by the Alfred P. Sloan Foundation, the Participating Institutions, the National Science Foundation,
 and the U.S. Department of Energy Office of Science. The SDSS-III web site is http://www.sdss3.org/.SDSS-III is managed by the Astrophysical Research Consortium for the Participating Institutions of the SDSS-III 
Collaboration including the University of Arizona, the Brazilian Participation Group, Brookhaven National 
Laboratory, University of Cambridge, Carnegie Mellon University, University of Florida, the French Participation 
Group, the German Participation Group, Harvard University, the Instituto de Astrofisica de Canarias, the Michigan 
State/Notre Dame/JINA Participation Group, Johns Hopkins University, Lawrence Berkeley National Laboratory, Max 
Planck Institute for Astrophysics, Max Planck Institute for Extraterrestrial Physics, New Mexico State University, 
New York University, Ohio State University, Pennsylvania State University, University of Portsmouth, Princeton 
University, the Spanish Participation Group, University of Tokyo, University of Utah, Vanderbilt University, 
University of Virginia, University of Washington, and Yale University.

\newpage


\begin{thebibliography}{}
  
  \bibitem{} Asher, D.J., Steel, D.I.: 1993, MNRAS 263, 179
  \bibitem{} Baudrand, A., et al.: 2001, A\&A 375, 275
  \bibitem{} Binzel, R. P.: 2000, P\&SS 48, 297
  \bibitem{} Boattini, A., et al.: 2001, A\&A 375, 293
  \bibitem{} Bonnarel, F., et al.: 2000, A\&AS 143, 33
  \bibitem{} Bottke, W.F., et al.: 2006, Annu. Rev. Earth Pl. Sc. 34, 157
  \bibitem{} Cardamone, C., et al.: 2009, MNRAS 399, 1191
  \bibitem{} Chapman, C.R.: 2004, E\&PSL 222, 1
  \bibitem{} Fischer, D.A. et al.: 2012, MNRAS 419, 2900
  \bibitem{} Fukugita, M., et al.: 1996, AJ 111, 1748
  \bibitem{} Hiroaki, A. et al.: 2011, ApJS 193, 29
  \bibitem{} Ivezi\'{c}, Ž., et al.: 2001, AJ 122, 2749
  \bibitem{} Juri\'{c}, M. et al.: 2002, AJ 124, 1776
  \bibitem{} Lawrence, A. et al.:2007, MNRAS 379, 1599
  \bibitem{} Mainzer, A. et al.: 2011, ApJ 743, 156
  \bibitem{} Marsden, B.G. et al.: 1992, IAUC 5670, 1
  \bibitem{} McNaught, R.H., Cass, C.P.: 1995, IAUC 6198, 1
  \bibitem{} Schwamb, M., et al: 2012, ApJ 754, 129
  \bibitem{} Pier, J.R., et al.: 2003, AJ 125, 1559
  \bibitem{} Vaduvescu, O. et al: 2011, P\&SS 59, 1632
  \bibitem{} York, D.G., et al.: 2000, AJ 120, 1579
  \bibitem{} Yoshida, F., Nakamura, T.: 2007, P\&SS 55, 1113
\end{thebibliography}
\end{document}